\batchmode
\makeatletter
\def\input@path{{/home/rgiordan/Documents/git_repos/VariationalRobustBayesPaper/NIPS2016/}}
\makeatother
\documentclass[english]{article}\usepackage[]{graphicx}\usepackage[]{color}
\makeatletter
\def\maxwidth{ %
  \ifdim\Gin@nat@width>\linewidth
    \linewidth
  \else
    \Gin@nat@width
  \fi
}
\makeatother

\definecolor{fgcolor}{rgb}{0.345, 0.345, 0.345}

\usepackage{framed}
\makeatletter
 {\par\unskip\endMakeFramed%
 \at@end@of@kframe}
\makeatother

\definecolor{shadecolor}{rgb}{.97, .97, .97}
\definecolor{messagecolor}{rgb}{0, 0, 0}
\definecolor{warningcolor}{rgb}{1, 0, 1}
\definecolor{errorcolor}{rgb}{1, 0, 0}
\newenvironment{knitrout}{}{} 

\usepackage{alltt}
\usepackage[T1]{fontenc}
\usepackage[latin9]{inputenc}
\usepackage{verbatim}
\usepackage{refstyle}
\usepackage{amsmath}
\usepackage{amsthm}
\usepackage{amssymb}
\usepackage[authoryear]{natbib}

\makeatletter

\AtBeginDocument{\providecommand\eqref[1]{\ref{eq:#1}}}
\AtBeginDocument{\providecommand\secref[1]{\ref{sec:#1}}}
\RS@ifundefined{subsecref}
  {\newref{subsec}{name = \RSsectxt}}
  {}
\RS@ifundefined{thmref}
  {\def\RSthmtxt{theorem~}\newref{thm}{name = \RSthmtxt}}
  {}
\RS@ifundefined{lemref}
  {\def\RSlemtxt{lemma~}\newref{lem}{name = \RSlemtxt}}
  {}

\usepackage[final]{nips_2016}

\setcitestyle{round}

\usepackage{hyperref}       
\usepackage{url}            
\usepackage{booktabs}       
\usepackage{amsfonts}       
\usepackage{nicefrac}       
\usepackage{microtype}      

\newcommand{\fig}[1]{Fig.~(\ref{fig:#1})}

\makeatother

\usepackage{babel}
\IfFileExists{upquote.sty}{\usepackage{upquote}}{}
\begin{document}

\global\long\def\constant{C}

\global\long\def\pthetapost{p_{\epsilon}^{x}}

\global\long\def\origpthetapost{p_{0}^{x}}

\global\long\def\qthetapost{q_{\epsilon}^{x}}

\global\long\def\origqthetapost{q_{0}^{x}}

\global\long\def\origprior{p_{0}}

\global\long\def\contamprior{p_{c}}

\global\long\def\mvprior{p_{mv}}

\global\long\def\prior{p\left(\theta\vert\epsilon\right)}

\global\long\def\etaopt{\eta^{*}}

\global\long\def\qthetapostarg{\qthetapost\left(\theta;\eta^{*}\right)}

\global\long\def\qtthetapostarg{\qthetapost\left(\theta;\eta^{*}\left(t\right)\right)}

\global\long\def\gtheta{g\left(\theta\right)}

\global\long\def\mbe{\mathbb{E}}

\global\long\def\mbeq{\mbe_{\qthetapost}}

\global\long\def\mbep{\mbe_{\pthetapost}}

\global\long\def\sens{S_{\epsilon}^{p_{c}}}

\global\long\def\origsens{S_{0}^{p_{c}}}

\global\long\def\qsens{\hat{S}_{\epsilon}^{p_{c}}}

\global\long\def\influence{I_{\epsilon}\left(\theta\right)}

\global\long\def\originfluence{I_{0}\left(\theta\right)}

\global\long\def\qinfluence{I_{\epsilon}^{q}\left(\theta\right)}

\global\long\def\origqinfluence{I_{0}^{q}\left(\theta\right)}

\global\long\def\epgtheta{\mbe_{\pthetapost}\left[\gtheta\right]}

\global\long\def\eqgtheta{\mbe_{\qthetapost}\left[\gtheta\right]}

\global\long\def\indep{\stackrel{indep}{\sim}}

\global\long\def\iid{\stackrel{iid}{\sim}}

\global\long\def\kl{\mathrm{KL}}

\global\long\def\argmin{\operatornamewithlimits{argmin}}

\global\long\def\argmax{\operatornamewithlimits{argmax}}

\title{The Prior Influence Function in Variational Bayes}

\author{   
Ryan Giordano \texttt{rgiordano@berkeley.edu}
\And
Tamara Broderick
\texttt{tbroderick@csail.mit.edu}
\And
Michael Jordan
\texttt{jordan@cs.berkeley.edu}
} 

\maketitle 
\begin{abstract}
In Bayesian analysis, the posterior follows from the data and a choice
of a prior and a likelihood. One hopes that the posterior is robust
to reasonable variation in the choice of prior, since this choice
is made by the modeler and is often somewhat subjective. A different,
equally subjectively plausible choice of prior may result in a substantially
different posterior, and so different conclusions drawn from the data.
Were this to be the case, our conclusions would not be robust to the
choice of prior. To determine whether our model is robust, we must
quantify how sensitive our posterior is to perturbations of our prior.

Variational Bayes (VB) methods are fast, approximate methods for posterior
inference. As with any Bayesian method, it is useful to evaluate the
robustness of a VB approximate posterior to changes in the prior.
In this paper, we derive VB versions of classical non-parametric local
robustness measures. In particular, we show that the influence function
of \citet{gustafson:2012:localrobustnessbook} has a simple, easy-to-calculate
closed form expression for VB approximations. We then demonstrate
how local robustness measures can be inadequate for non-local prior
changes, such as replacing one prior entirely with another. We propose
a simple approximate non-local robustness measure and demonstrate
its effectiveness on a simulated data set.
\end{abstract}

\section{Local robustness and the influence function\label{sec:local_robustness}}

Bayesian robustness studies how changes to the model (i.e., the prior
and likelihood) and to the data affect the posterior. If important
aspects of the posterior are meaningfully sensitive to subjectively
reasonable perturbations of the inputs, then the posterior is ``non-robust''
to these perturbations. In this paper, we focus on quantifying the
sensitivity of posterior means to perturbations of the prior \textendash{}
either infinitesimally mixing or completely replacing the original
prior with another ``contaminating prior''. Our methods allow fast
estimation of sensitivity to any contaminating prior without re-fitting
the model. We follow and extend the work of \citet{gustafson:1996:localposterior}
and \citet{gustafson:2012:localrobustnessbook} to variational Bayes
and to approximate non-local measures of sensitivity. For a more general
review of Bayesian robustness, see \citet{berger:2012:robust}.

We will now define some terminology. Denote our $N$ data points by
$x=\left(x_{1},\ldots,x_{N}\right)$ with $x_{n}\in\mathbb{R}^{D}$.
Denote our parameter by the vector $\theta\in\mathbb{R}^{K}$. We
will suppose that we are interested in the robustness of our prior
to a scalar parameter $\epsilon$ where our prior can be written as
$\prior$. Let $\pthetapost$ denote the posterior distribution of
$\theta$ with prior given by $\epsilon$ and conditional on $x$,
as given by Bayes' Theorem: $\pthetapost\left(\theta\right):=p\left(\theta\vert x,\epsilon\right)=\frac{p\left(x\vert\theta\right)\prior}{p\left(x\right)}$.

A typical end product of a Bayesian analysis might be a posterior
expectation of some function, $\epgtheta$, which is a functional
of $g\left(\theta\right)$ and $\pthetapost\left(\theta\right)$.
Local robustness considers how much $\epgtheta$ changes locally in
response to small perturbations in the value of $\epsilon$ \citep{gustafson:2012:localrobustnessbook}.
In the present work, we consider mixing our original prior, $\origprior\left(\theta\right)$,
with some known alternative functional form, $\contamprior\left(\theta\right)$:
\begin{eqnarray}
\prior & = & \left(1-\epsilon\right)\origprior\left(\theta\right)+\epsilon\contamprior\left(\theta\right)\textrm{ for }\epsilon\in\left[0,1\right].\label{eq:epsilon_contamination}
\end{eqnarray}
This is known as epsilon contamination (the subscript $c$ stands
for ``contamination''), and its construction guarantees that the
perturbed prior is properly normalized. The contaminating prior, $\contamprior\left(\theta\right)$
need not be in the same parametric family as $\origprior\left(\theta\right)$,
so as $\contamprior\left(\theta\right)$ ranges over all possible
priors, \eqref{epsilon_contamination} represents an expressive class
of perturbations. Under mild assumptions (given in \secref{app_epsilon_sensitivity}),
the local sensitivity measure at the prior $\prior$ given by a particular
$\epsilon$ is 
\begin{eqnarray}
\sens & = & \left.\frac{d\epgtheta}{d\epsilon}\right|_{\epsilon}=\textrm{Cov}_{\pthetapost}\left(g\left(\theta\right),\frac{\contamprior\left(\theta\right)-\origprior\left(\theta\right)}{\origprior\left(\theta\right)+\epsilon\left(\contamprior\left(\theta\right)-\origprior\left(\theta\right)\right)}\right).\label{eq:local_robustness}
\end{eqnarray}

The definition in \eqref{local_robustness} depends on a choice of
$p_{c}\left(\theta\right)$, which we denote with a superscript on
$\sens$. At $\epsilon=0$, we recover the local sensitivity around
$\origprior\left(\theta\right)$, which we denote $\origsens$. 

Rather than choose some finite set of $\contamprior\left(\theta\right)$
and calculate their corresponding $\sens$, one can work with a single
function that summarizes the effect of any $\contamprior\left(\theta\right)$,
called the ``influence function'' \citep{gustafson:2012:localrobustnessbook}.
Observing that \eqref{local_robustness} is a linear functional of
$\contamprior\left(\theta\right)$ when $g\left(\theta\right)$, $\epsilon$,
and $\prior$ are fixed, the influence function (when it exists) is
defined as the linear operator $\influence$ that characterizes the
dependence of $\sens$ on $\contamprior\left(\theta\right)$: 
\begin{eqnarray}
\sens & = & \int\influence\contamprior\left(\theta\right)d\theta\quad\textrm{where}\quad\influence:=\frac{\pthetapost\left(\theta\right)}{\prior}\left(g\left(\theta\right)-\epgtheta\right).\label{eq:influence_definition}
\end{eqnarray}
At $\epsilon=0$, we recover the local sensitivity around $\origprior\left(\theta\right)$,
which we denote $\originfluence$. When perturbing a low-dimensional
marginal of the prior, $\originfluence$ is an easy-to-visualize summary
of the effect of sensitivity to an arbitrary $\contamprior\left(\theta\right)$
using quantities calculated only under $\origprior\left(\theta\right)$
(see the example in \secref{Experiments} and the extended discussion
in \citet{gustafson:2012:localrobustnessbook}). Additionally, the
worst case prior in a suitably defined metric ball around $\origprior\left(\theta\right)$
is a functional of the influence function, as shown in \citet{gustafson:2012:localrobustnessbook}. 

\section{Variational approximation and linear response\label{sec:Variational-approximation}}

We now derive a version of \eqref{local_robustness} for Variational
Bayes (VB) approximations to the posterior. Recall that an variational
approximate posterior is a distribution selected to minimize the Kullback-Liebler
(KL) divergence to $\pthetapost$ across distributions $q$ in some
class $\mathcal{Q}$. Let $\qthetapost$ denote the variational approximation
to posterior $\pthetapost$. We assume that distributions in $\mathcal{Q}$
are smoothly parameterized by a finite-dimensional parameter $\eta$
whose optimum lies in the interior of some feasible set $\Omega_{\eta}$.

We would like to calculate the local robustness measures of \secref{local_robustness}
for the variational approximation $\qthetapost$, but a direct evaluation
of the covariance in \eqref{local_robustness} can be misleading.
For example, a common choice of the approximating family$\mathcal{Q}$
is the class of distributions that factorize across $\theta$. This
is known as the ``mean field approximation'' \citep{wainwright2008graphical}.
By construction, a mean field approximation does not model covariances
between independent components of $\theta$, so a naive estimate of
the covariance in \eqref{local_robustness} may erroneously suggest
that the prior on one component of $\theta$ cannot affect the posterior
on another.

However, for VB approximations, we can evaluate the derivative on
the left hand side of \eqref{local_robustness} directly. Using linear
response variational Bayes (LRVB) \citep{giordano:2016:robust,giordano:2015:lrvb},
we have

\begin{eqnarray}
\left.\frac{d}{d\epsilon}\mbeq\left[g\left(\theta\right)\right]\right|_{\epsilon} & = & \int\frac{\qthetapost\left(\theta\right)}{\prior}q_{\eta}\left(\theta\right)^{T}H^{-1}g_{\eta}p_{c}\left(\theta\right)d\theta\label{eq:lrvb_epsilon_sensitivity}\\
\textrm{where }g_{\eta}:=\frac{\partial\mbeq\left[g\left(\theta\right)\right]}{\partial\eta}\textrm{, }\quad q_{\eta}\left(\theta\right) & := & \frac{\partial\log\qthetapost\left(\theta;\eta\right)}{\partial\eta}\textrm{, and }\quad H:=\frac{\partial^{2}KL\left(\qthetapost\left(\theta;\eta\right)||\pthetapost\right)}{\partial\eta\partial\eta^{T}}.\nonumber 
\end{eqnarray}

It follows immediately from the definition in \eqref{influence_definition}
that we can define the \textit{variational influence function}
\begin{align}
\qinfluence & :=\frac{\qthetapost\left(\theta\right)}{\prior}q_{\eta}\left(\theta\right)^{T}H^{-1}g_{\eta}\label{eq:lrvb_influence_function}
\end{align}

that captures the sensitivity of $\mbeq\left[g\left(\theta\right)\right]$
just as $\influence$ captures the sensitivity of $\mbep\left[g\left(\theta\right)\right]$. 

The VB versions of epsilon sensitivity measures have some advantages
and disadvantages relative to using Markov Chain Monte Carlo (MCMC)
to evaluate the exact sensitivities in \secref{local_robustness}.
Using MCMC samples from $\origpthetapost$, one can form a Monte Carlo
estimate of the covariance in \eqref{local_robustness}, though the
sample variance may be infinite when $\contamprior$ has heavier tails
than $\origprior$. The extent to which this is a real problem in
practice will vary. Similarly, one must take care in numerically evaluating
\eqref{lrvb_epsilon_sensitivity}, since naively sampling from $\qthetapost$
may also result in infinite variance due to the term $\prior$ in
the denominator. Since we have a closed form for $\qthetapost$, we
can instead evaluate \eqref{lrvb_epsilon_sensitivity} as an integral
over $\contamprior$ using importance sampling, as described in \secref{app_Importance-sampling}.
Still, providing efficient estimates of \eqref{lrvb_epsilon_sensitivity}
for high-dimensional, non-conjugate, heavy-tailed $\contamprior$
remains a challenge. Finally, in contrast to \eqref{influence_definition},
where we do not generally have a closed form expression for $\pthetapost$,
every term in \eqref{lrvb_influence_function} is known. This means
it is easier to evaluate the influence function for VB approximations
than from MCMC draws, especially far from the posterior.

\section{Non-local approximation}

Equation (\ref{eq:influence_definition}) quantifies the effect of
adding an infinitesimal amount of the contaminating prior. In practice,
we may also want to evaluate intermediate values of $\epsilon$, particularly
$\epsilon=1$, which represents completely replacing $\origprior\left(\theta\right)$
with $p_{c}\left(\theta\right)$. Since $p_{c}\left(\theta\right)$
may be quite different from $\origprior\left(\theta\right)$, this
is a non-local robustness measure. For MCMC samples, one can use importance
sampling, which is essentially equivalent to evaluating the covariance
in \eqref{local_robustness} (with the same problem of infinite variance
\textendash{} see \secref{app_mcmc_importance_sampling}). For VB,
however, we either need to re-fit the model for each new prior (which
may be time consuming) or somehow use the local information at $\origqthetapost$.
In this paper, we investigate the latter. For the remainder of this
section, since our results are general, we will discuss using local
information in $\origpthetapost$. However, the reader should keep
in mind that the ultimate goal is to apply the insights gained to
the variational approximation $\origqthetapost$.

One might hope to linearly extrapolate from $\epsilon=0$ to $\epsilon=1$
using the slope $\origsens$ at $\epsilon=0$. That is, we might hope
that $\left.\epgtheta\right|_{\epsilon=0}^{\epsilon=1}\approx\left(1-0\right)\left.\frac{d\epgtheta}{d\epsilon}\right|_{\epsilon=0}$.
However, as we will now show, this is not realistic when one of the
two priors is more consistent with the data than the other. Inspection
of \eqref{influence_definition} shows that posterior expectations
are highly sensitive to perturbations of priors which are inconsistent
with the data: if $\prior$ is small in an area of the $\theta$ space
where $\pthetapost$ is not small, then the influence function $\influence$
will be quite large. The model will have high sensitivity to any contaminating
prior, $p_{c}\left(\theta\right)$, that is more consistent with the
model than $\prior$ at $\epsilon$. In particular, this is true at
$\epsilon=0$ where $\prior=\origprior\left(\theta\right)$ if $\origprior\left(\theta\right)$
is inconsistent with the data. In fact, as we show in \secref{app_epsilon_sensitivity},
\begin{align}
\left.\epgtheta\right|_{\epsilon=0}^{\epsilon=1} & =\int_{0}^{1}\left.\frac{d\epgtheta}{d\epsilon}\right|_{\epsilon}d\epsilon=\frac{\int p\left(x\vert\theta\right)\contamprior\left(\theta\right)d\theta}{\int p\left(x\vert\theta\right)\origprior\left(\theta\right)d\theta}\origsens.\label{eq:global_local_sens_evidence}
\end{align}
When the model evidence is very different for $\contamprior$ and
$\origprior$, e.g. when $\int p\left(x\vert\theta\right)\origprior\left(\theta\right)d\theta\ll\int p\left(x\vert\theta\right)\contamprior\left(\theta\right)d\theta$
as in \secref{Experiments}, the extrapolated slope $\origsens$ can
be quite different from the effect of replacing completely replacing
$\origprior\left(\theta\right)$ with $p_{c}\left(\theta\right)$. 

However, as $\epsilon$ grows away from zero and the new prior $p_{c}\left(\theta\right)$
is taken into account, the influence function will shrink. Observe
that as a function of $\epsilon$ \eqref{local_robustness}, one can
show (see \secref{app_epsilon_sensitivity}) that 

\begin{align}
\left.\frac{d\epgtheta}{d\epsilon}\right|_{\epsilon} & \le\textrm{max}\left\{ \frac{1}{\epsilon},\frac{1}{1-\epsilon}\right\} \mbep\left[\left|g\left(\theta\right)-\epgtheta\right|\right].\label{eq:sensitivity_bound}
\end{align}

For $\epsilon=0$ or $1$, this bound is vacuous, since the ratio
$\contamprior\left(\theta\right)/\origprior\left(\theta\right)$ can
be arbitrarily large in areas assigned positive probability by $\pthetapost$.
However, for intermediate values of $\epsilon$, such as $\frac{1}{2}$,
the bound is quite strong. In other words, contamination with $p_{c}\left(\theta\right)$
can have great influence on $\epgtheta$ when $\epsilon$ is close
to the boundaries of $\left[0,1\right]$, but once $\contamprior\left(\theta\right)$
is taken into account with intermediate $\epsilon$, its influence
is tightly bounded by \eqref{sensitivity_bound}. In this sense, the
value of $\left.\frac{d\epgtheta}{d\epsilon}\right|_{\epsilon}$ is
most atypical of its value across the interval $\epsilon\in\left[0,1\right]$
at its endpoints. A real-life example of exactly this phenomenon is
shown in \secref{Experiments} in \fig{MicrocreditInfluenceFunction}. 

This suggests replacing the derivative at $\epsilon=0$ with an average
of the derivative over the interval $\epsilon\in\left[0,1\right]$.
To do this, note that the difficulty of the integral in \eqref{global_local_sens_evidence}
is the complicated dependence of $\pthetapost$ on $\epsilon$ in
$\influence$. However, we can approximate the integral by keeping
$\pthetapost$ fixed at $\origpthetapost$ so that $\influence$ only
depends on $\epsilon$ through $\prior$. Under this approximation,
the integral can be evaluated analytically (see \secref{app_Mean-Value}),
giving the contaminating pseudo-density, $\mvprior\left(\theta\right)$,
which represents the approximate effect of integrating over $\epsilon$
from $0$ to $1$:
\begin{align}
\left.\epgtheta\right|_{\epsilon=0}^{\epsilon=1} & \approx\int\originfluence p_{mv}\left(\theta\right)d\theta\quad\quad p_{mv}\left(\theta\right):=\frac{p_{c}\left(\theta\right)\origprior\left(\theta\right)}{p_{c}\left(\theta\right)-\origprior\left(\theta\right)}\log\frac{\contamprior\left(\theta\right)}{\origprior\left(\theta\right)}.\label{eq:mean_value_definition}
\end{align}

In the notation $\mvprior\left(\theta\right)$, ``mv'' stands for
``mean value'', by analogy with the mean value theorem for functions
of real numbers. As shown in \secref{Experiments}, using $\mvprior\left(\theta\right)$
with $\originfluence$ rather than $\contamprior\left(\theta\right)$
can represent a significant improvement over \eqref{influence_definition}
in practice when extrapolating to $\epsilon=1$. We will focus on
using $\mvprior$ with the variational approximations described in
\secref{Variational-approximation}.

\section{Experiments\label{sec:Experiments}}

\newcommand{\mcPriorEta}{15.010}
\newcommand{\mcPriorMuInfo}{0.111}
\newcommand{\mcPriorScaleAlpha}{20.010}
\newcommand{\mcPriorScaleBeta}{20.010}
\newcommand{\mcPriorTauAlpha}{2.010}
\newcommand{\mcPriorTauBeta}{2.010}
\newcommand{\numGroups}{30}
\newcommand{\numObs}{3000}
\newcommand{\mcPriorStudentTDF}{1}

We demonstrate our methods using simulated data from a hierarchical
model described in \secref{app_Microcredit-model}. Here, we will
demonstrate that our sensitivity measures accurately predict the changes
in VB solutions. We discuss the close match between the VB and MCMC
results in \secref{app_Microcredit-model}. The results below are
for the sensitivity of the expectation $\mbe_{\origpthetapost}\left[\mu_{11}\right]$
to the prior $p\left(\mu\right)$, though similar results are easily
calculated for any other low-dimensional prior marginal or posterior
expectation.

We generate data using true parameters that are far from $\origprior\left(\theta\right)$
so that our model is not robust to perturbations, as can be seen by
the large values of the influence function, which is pictured in the
top left panel of \fig{MicrocreditInfluenceFunction}. The posterior
mean is shown with a black dot, indicating that, though large, the
influence function is very highly concentrated around the posterior
mean. The top right panel of \fig{MicrocreditInfluenceFunction} indicates
how $\mbeq\left[\mu_{11}\right]$ depends on $\epsilon$ near $\epsilon=0$.
The slope is very steep at $\epsilon=0$, reflecting the fact that
$\contamprior\left(\theta\right)$ takes values much larger than $\origprior\left(\theta\right)$
near the posterior where the influence function is very high. However,
it very quickly levels off for $\epsilon$ only slightly above zero.

The top right panel of \fig{MicrocreditInfluenceFunction} indicates
that extrapolating from the slope $\origsens$  at $\epsilon=0$ will
radically over-estimate the effect of replacing $\origprior\left(\theta\right)$
with $\contamprior\left(\theta\right)$. This is confirmed in the
bottom left panel, which has the actual change in $\mbeq\left[\theta\right]$
on the x-axis and $\int\qinfluence\contamprior\left(\theta\right)d\theta$
on the y-axis. Clearly, the extrapolation is unrealistic. However,
the right panel of \fig{MicrocreditInfluenceFunction} demonstrates
that $\int\qinfluence p_{mv}\left(\theta\right)d\theta$ accurately
matches the effect of replacing $\origprior\left(\theta\right)$ with
$\contamprior\left(\theta\right)$. Note the different ranges in the
y-axis (which prohibit plotting the two graphs on the same scale).
The error bars represent importance sampling error.

\begin{knitrout}
\definecolor{shadecolor}{rgb}{0.969, 0.969, 0.969}\color{fgcolor}\begin{figure}[ht!]

{\centering \includegraphics[width=0.49\linewidth,height=0.196\linewidth]{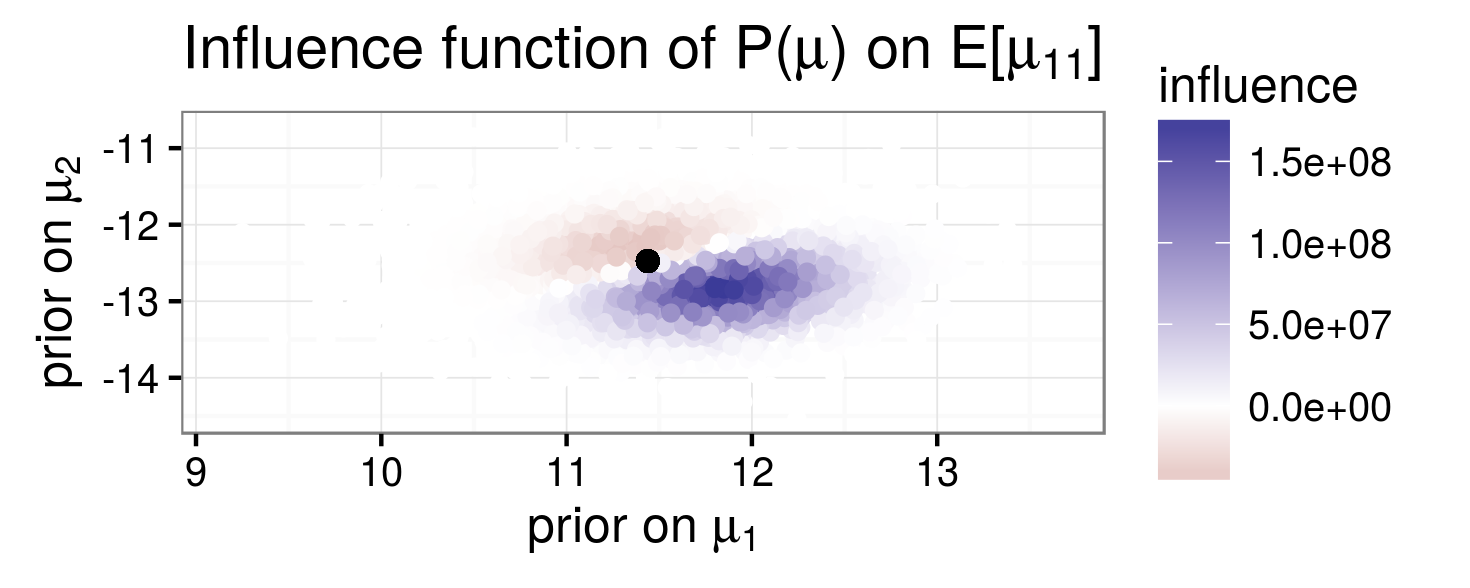} 
\includegraphics[width=0.49\linewidth,height=0.196\linewidth]{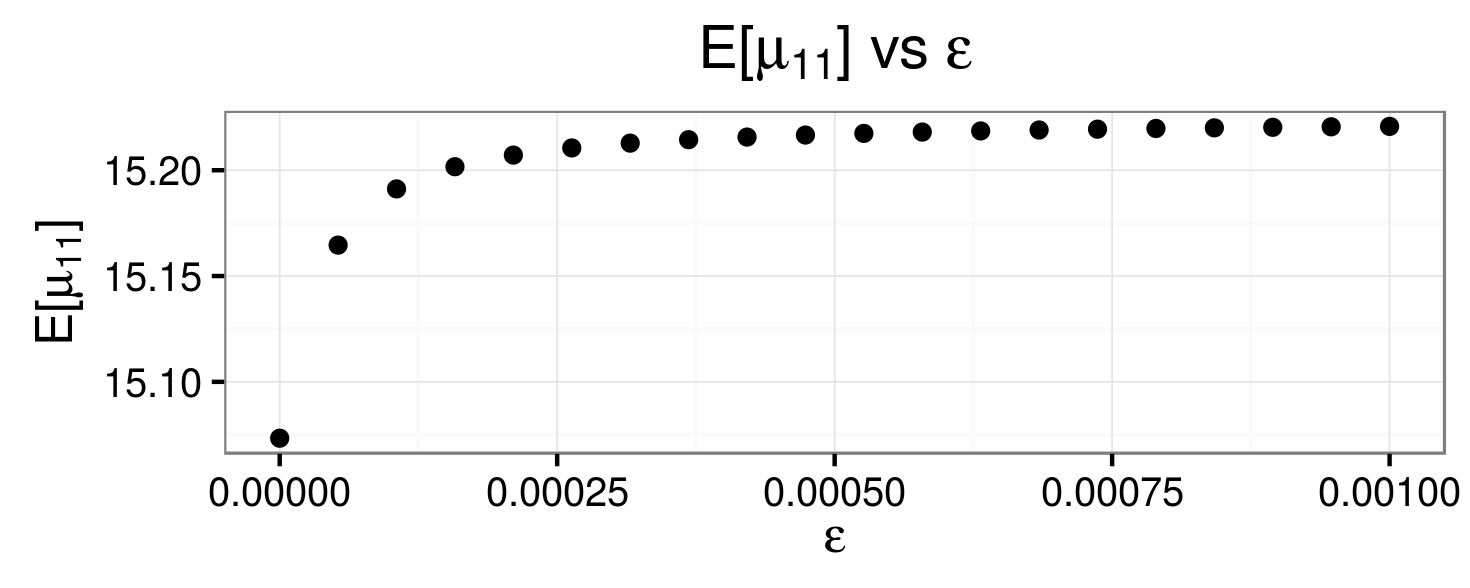} 
\includegraphics[width=0.49\linewidth,height=0.196\linewidth]{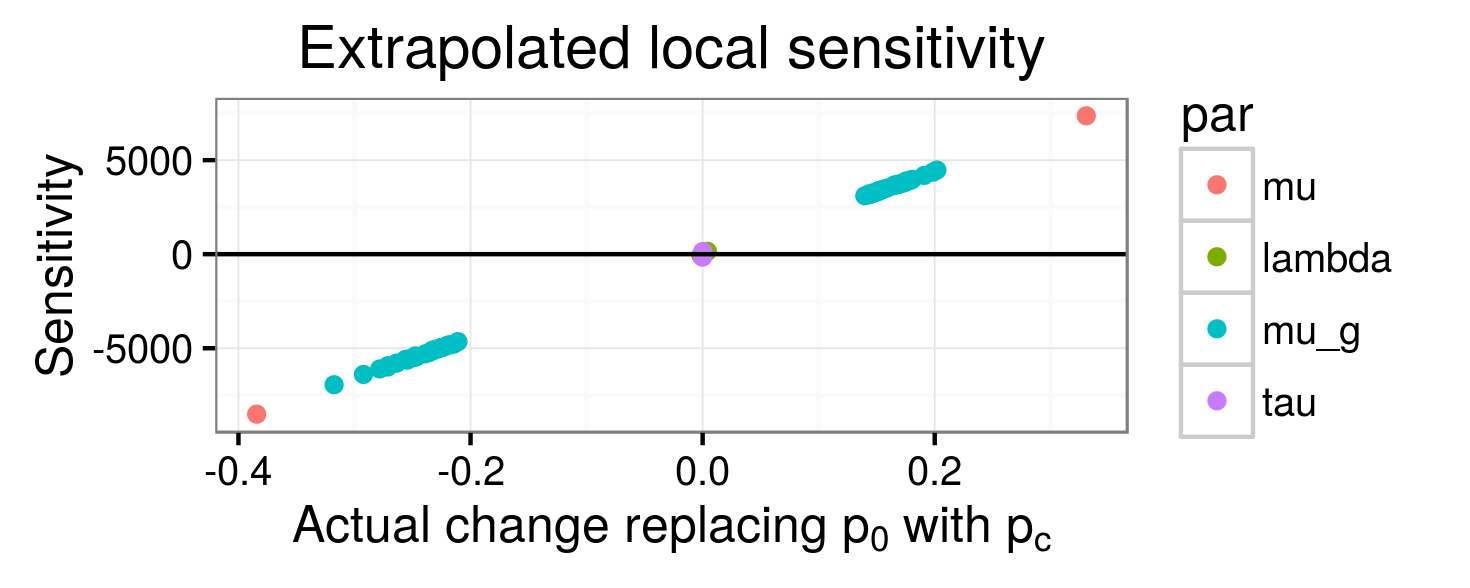} 
\includegraphics[width=0.49\linewidth,height=0.196\linewidth]{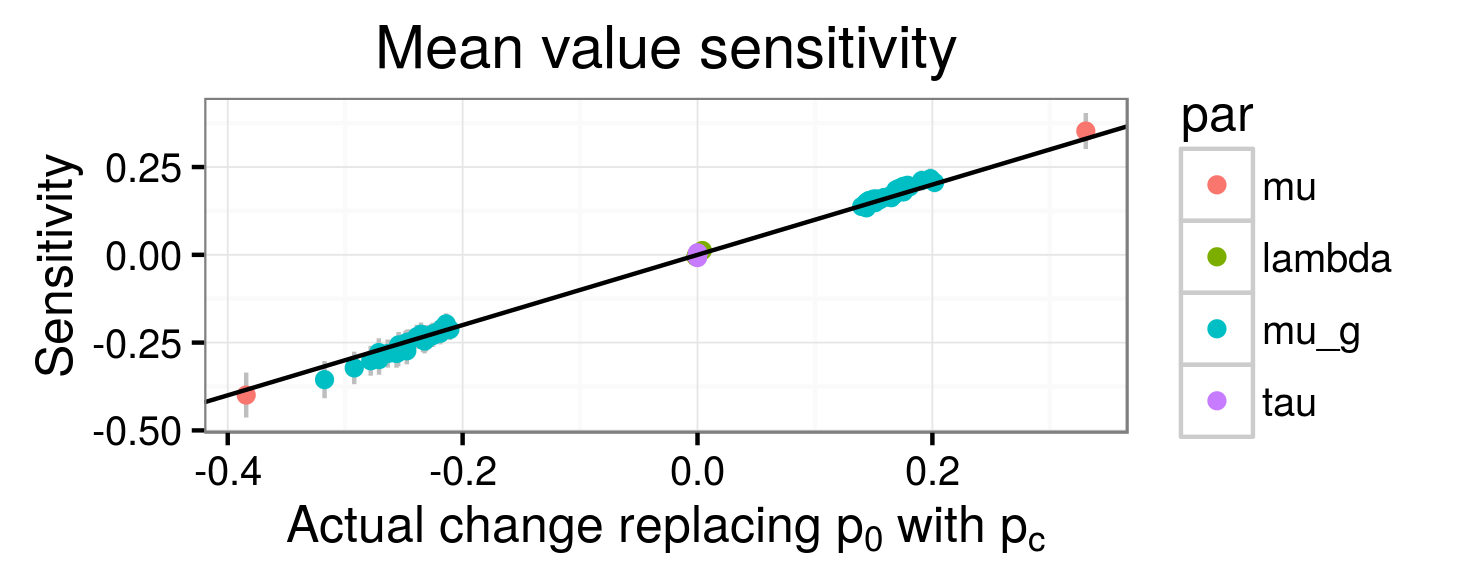} 

}

\caption[Simulation results]{Simulation results}\label{fig:MicrocreditInfluenceFunction}
\end{figure}

\end{knitrout}

\paragraph*{{\footnotesize{}Acknowledgements}}

{\footnotesize{}The paper was greatly improved through the insightful
comments of our anonymous reviewers. Ryan Giordano was supported by
the Director, Office of Science, Office of Advanced Scientific Computing
Research, Applied Mathematics program of the U.S. Department of Energy
under Contract No. DE-AC02-05CH1123.}{\footnotesize \par}

\clearpage{}

\bibliographystyle{plainnat}
\bibliography{variational_robustness}

\clearpage{}

\appendix

\part*{Appendices}

\section{Epsilon sensitivity\label{sec:app_epsilon_sensitivity}}

Throughout the paper, we make the following assumptions:
\begin{itemize}
\item Assumption 1. For all $\epsilon\in\left[0,1\right]$, $\prior$ is
strictly positive where $\theta$ has positive measure.
\item Assumption 2. Both $p\left(\theta\vert x\right)$ and $g\left(\theta\right)p\left(\theta\vert x\right)$
are bounded as a function of $\theta$ .
\end{itemize}
Under these assumptions, \eqref{local_robustness} follows directly
from \citet{gustafson:1996:localposterior} result 8. For completeness,
we reproduce a slightly simpler proof under the slightly less clearly-articulated
assumption that we can exchange integration and differentiation as
described, for example, in \citet[Appendix A.5]{durrett:2010:probability}.
Denote by $\prior$ any distribution parameterized by the scalar $\epsilon$
(not necessarily a prior). Then by direct differentiation, 
\begin{align}
\frac{d\mbe_{\prior}\left[g\left(\theta\right)\right]}{d\epsilon} & =\frac{d}{d\epsilon}\int\prior g\left(\theta\right)d\theta\nonumber \\
 & =\int\frac{d\prior}{d\epsilon}g\left(\theta\right)d\theta\nonumber \\
 & =\int\frac{d\log\prior}{d\epsilon}\prior g\left(\theta\right)d\theta.\label{eq:differentiate_under_integral}
\end{align}

By applying \eqref{differentiate_under_integral} to $g\left(\theta\right)=1$,
we see that $\mbe_{\prior}\left[\frac{d\log\prior}{d\epsilon}\right]=\int\frac{d\log\prior}{d\epsilon}\prior d\theta=0$,
so we can subtract $0=\mbe_{\prior}\left[g\left(\theta\right)\right]\mbe_{\prior}\left[\frac{d\log\prior}{d\epsilon}\right]$
to get
\begin{align}
\frac{d\mbe_{\prior}\left[g\left(\theta\right)\right]}{d\epsilon} & =\textrm{Cov}_{\prior}\left(g\left(\theta\right),\frac{d\log\prior}{d\epsilon}\right).\label{eq:covariance_sensitivity}
\end{align}

To derive \eqref{local_robustness}, we simply observe that
\begin{align*}
\frac{d\log\prior}{d\epsilon} & =\frac{d}{d\epsilon}\log\left(\left(1-\epsilon\right)\origprior\left(\theta\right)+\epsilon\contamprior\left(\theta\right)\right)\\
 & =\frac{\contamprior\left(\theta\right)-\origprior\left(\theta\right)}{\origprior\left(\theta\right)+\epsilon\left(\contamprior\left(\theta\right)-\origprior\left(\theta\right)\right)}.
\end{align*}

Note that the assumptions also suffice to assure that the covariance
is bounded.

Next, we observe the simple relationship between epsilon sensitivity
at $\epsilon=0$ and the effect of replacing one prior with another.
First, defining the normalizing constants
\begin{align*}
C_{0} & :=\int p\left(x\vert\theta\right)\origprior\left(\theta\right)d\theta\\
C_{1} & :=\int p\left(x\vert\theta\right)\contamprior\left(\theta\right)d\theta\\
\mbe_{\origpthetapost}\left[\frac{\contamprior\left(\theta\right)}{\origprior\left(\theta\right)}\right] & =\int\frac{p\left(x\vert\theta\right)\origprior\left(\theta\right)}{C_{0}}\frac{\contamprior\left(\theta\right)}{\origprior\left(\theta\right)}d\theta=\frac{C_{1}}{C_{0}},
\end{align*}

by straightforward manipulation we have
\begin{align*}
\left.\mbe_{\pthetapost\left(\theta\right)}\left[g\left(\theta\right)\right]\right|_{\epsilon=1}-\left.\mbe_{\pthetapost\left(\theta\right)}\left[g\left(\theta\right)\right]\right|_{\epsilon=0} & =\int\frac{p\left(x\vert\theta\right)\contamprior\left(\theta\right)}{C_{1}}g\left(\theta\right)d\theta-\mbe_{\origpthetapost\left(\theta\right)}\left[g\left(\theta\right)\right]\\
 & =\int\frac{p\left(x\vert\theta\right)\contamprior\left(\theta\right)}{C_{1}}\frac{\origprior\left(\theta\right)}{C_{0}}\frac{C_{0}}{\origprior\left(\theta\right)}g\left(\theta\right)d\theta-\frac{C_{0}}{C_{1}}\mbe_{\origpthetapost}\left[\frac{\contamprior\left(\theta\right)}{\origprior\left(\theta\right)}\right]\mbe_{\origpthetapost\left(\theta\right)}\left[g\left(\theta\right)\right]\\
 & =\frac{C_{0}}{C_{1}}\left(\mbe_{\origpthetapost\left(\theta\right)}\left[\frac{\contamprior\left(\theta\right)}{\origprior\left(\theta\right)}g\left(\theta\right)\right]-\mbe_{\origpthetapost}\left[\frac{\contamprior\left(\theta\right)}{\origprior\left(\theta\right)}\right]\mbe_{\origpthetapost\left(\theta\right)}\left[g\left(\theta\right)\right]\right)\\
 & =\frac{C_{0}}{C_{1}}\textrm{Cov}_{\origpthetapost\left(\theta\right)}\left(g\left(\theta\right),\frac{\contamprior\left(\theta\right)}{\origprior\left(\theta\right)}\right)\\
 & =\frac{C_{0}}{C_{1}}\left.\frac{d\epgtheta}{d\epsilon}\right|_{\epsilon=0}.
\end{align*}

Finally, we derive \eqref{sensitivity_bound} using \eqref{local_robustness}:

\begin{align*}
\left|\left.\frac{d\epgtheta}{d\epsilon}\right|_{\epsilon}\right| & =\left|\textrm{Cov}_{\pthetapost}\left(g\left(\theta\right),\frac{\contamprior\left(\theta\right)-\origprior\left(\theta\right)}{\origprior\left(\theta\right)+\epsilon\left(\contamprior\left(\theta\right)-\origprior\left(\theta\right)\right)}\right)\right|\\
 & \le\mbe_{\pthetapost}\left[\left|g\left(\theta\right)-\epgtheta\right|\right]B_{\epsilon}
\end{align*}

where
\begin{align*}
B_{\epsilon} & =\sup_{\theta}\left|\frac{\contamprior\left(\theta\right)-\origprior\left(\theta\right)}{\origprior\left(\theta\right)+\epsilon\left(\contamprior\left(\theta\right)-\origprior\left(\theta\right)\right)}\right|\\
 & =\sup_{\theta}\begin{cases}
\frac{\contamprior\left(\theta\right)-\origprior\left(\theta\right)}{\origprior\left(\theta\right)+\epsilon\left(\contamprior\left(\theta\right)-\origprior\left(\theta\right)\right)} & \textrm{when }\contamprior\left(\theta\right)>\origprior\left(\theta\right)\\
\frac{\origprior\left(\theta\right)-\contamprior\left(\theta\right)}{\origprior\left(\theta\right)+\epsilon\left(\contamprior\left(\theta\right)-\origprior\left(\theta\right)\right)} & \textrm{when }\contamprior\left(\theta\right)\le\origprior\left(\theta\right)
\end{cases}\\
 & =\sup_{\theta}\begin{cases}
\frac{\contamprior\left(\theta\right)-\origprior\left(\theta\right)}{\origprior\left(\theta\right)+\epsilon\left(\contamprior\left(\theta\right)-\origprior\left(\theta\right)\right)} & \textrm{when }\contamprior\left(\theta\right)>\origprior\left(\theta\right)\\
\frac{\origprior\left(\theta\right)-\contamprior\left(\theta\right)}{\contamprior\left(\theta\right)+\left(1-\epsilon\right)\left(\origprior\left(\theta\right)-\contamprior\left(\theta\right)\right)} & \textrm{when }\contamprior\left(\theta\right)\le\origprior\left(\theta\right)
\end{cases}\\
 & \le\sup_{\theta}\begin{cases}
\frac{\contamprior\left(\theta\right)-\origprior\left(\theta\right)}{\epsilon\left(\contamprior\left(\theta\right)-\origprior\left(\theta\right)\right)} & \textrm{when }\contamprior\left(\theta\right)>\origprior\left(\theta\right)\\
\frac{\origprior\left(\theta\right)-\contamprior\left(\theta\right)}{\left(1-\epsilon\right)\left(\origprior\left(\theta\right)-\contamprior\left(\theta\right)\right)} & \textrm{when }\contamprior\left(\theta\right)\le\origprior\left(\theta\right)
\end{cases}\\
 & \le\textrm{max}\left\{ \frac{1}{\epsilon},\frac{1}{1-\epsilon}\right\} .
\end{align*}

\section{Mean Value Contaminating Prior\label{sec:app_Mean-Value}}

Under the assumption that $\pthetapost\approx\origpthetapost$,

\begin{align*}
\left.\epgtheta\right|_{\epsilon=1}-\left.\epgtheta\right|_{\epsilon=0} & =\int_{0}^{1}\int\frac{\pthetapost\left(\theta\right)\contamprior\left(\theta\right)}{\prior}\left(g\left(\theta\right)-\mbe_{\left.\pthetapost\left(\theta\right)\right|_{\epsilon=0}}\left[g\left(\theta\right)\right]\right)d\theta d\epsilon\\
 & \approx\int_{0}^{1}\int\frac{\origpthetapost\left(\theta\right)\contamprior\left(\theta\right)}{\prior}\left(g\left(\theta\right)-\mbe_{\left.\pthetapost\left(\theta\right)\right|_{\epsilon=0}}\left[g\left(\theta\right)\right]\right)d\theta d\epsilon\\
 & =\int\origpthetapost\left(\theta\right)\left(g\left(\theta\right)-\mbe_{\origpthetapost\left(\theta\right)}\left[g\left(\theta\right)\right]\right)\int_{0}^{1}\frac{\contamprior\left(\theta\right)}{\prior}d\epsilon d\theta\\
 & =\int\origpthetapost\left(\theta\right)\left(g\left(\theta\right)-\mbe_{\origpthetapost\left(\theta\right)}\left[g\left(\theta\right)\right]\right)\frac{p_{c}\left(\theta\right)}{p_{c}\left(\theta\right)-\origprior\left(\theta\right)}\left(\log\contamprior\left(\theta\right)-\log\origprior\left(\theta\right)\right)d\theta\\
 & =\int\originfluence\frac{p_{c}\left(\theta\right)\origprior\left(\theta\right)}{p_{c}\left(\theta\right)-\origprior\left(\theta\right)}\left(\log\contamprior\left(\theta\right)-\log\origprior\left(\theta\right)\right)d\theta.
\end{align*}

Where we have used
\begin{align*}
\int_{0}^{1}\frac{\contamprior\left(\theta\right)}{\prior}d\epsilon & =\int_{0}^{1}\frac{\contamprior\left(\theta\right)}{\left(1-\epsilon\right)\origprior\left(\theta\right)+\epsilon p_{c}\left(\theta\right)}d\epsilon\\
 & =\frac{p_{c}\left(\theta\right)}{p_{c}\left(\theta\right)-\origprior\left(\theta\right)}\int_{0}^{1}\frac{d}{d\epsilon}\log\left(\left(1-\epsilon\right)\origprior\left(\theta\right)+\epsilon p_{c}\left(\theta\right)\right)d\epsilon\\
 & =\frac{p_{c}\left(\theta\right)}{p_{c}\left(\theta\right)-\origprior\left(\theta\right)}\left(\log\contamprior\left(\theta\right)-\log\origprior\left(\theta\right)\right).
\end{align*}

Consequently, applying \eqref{influence_definition} with the pseudo-density
\begin{align*}
p_{mv} & :=\frac{p_{c}\left(\theta\right)\origprior\left(\theta\right)}{p_{c}\left(\theta\right)-\origprior\left(\theta\right)}\left(\log\contamprior\left(\theta\right)-\log\origprior\left(\theta\right)\right)
\end{align*}

represents an approximation to the quantity $\left.\epgtheta\right|_{\epsilon=1}-\left.\epgtheta\right|_{\epsilon=0}$,
which is the effect of completely replacing $\origprior\left(\theta\right)$
with $\contamprior\left(\theta\right)$.

\section{Comparison with MCMC importance sampling\label{sec:app_mcmc_importance_sampling}}

In this section, we show that using importance sampling with MCMC
samples to calculate the local sensitivity \eqref{local_robustness}
is precisely equivalent to using the same MCMC samples to estimate
the covariance in \eqref{covariance_sensitivity} directly. Suppose,
without loss of generality, we have samples $\theta_{i}$ drawn from
$\origpthetapost\left(\theta\right)$ 
\begin{align*}
\theta_{i} & \sim\origpthetapost\left(\theta\right)\\
\mbe_{\origpthetapost}\left[g\left(\theta\right)\right] & \approx\sum_{i}g\left(\theta_{i}\right).
\end{align*}

If we could calculate the normalizing constants, the importance sampling
estimate for $\mbe_{\pthetapost}\left[g\left(\theta\right)\right]$
would be
\begin{align*}
\mbe_{\pthetapost}\left[g\left(\theta\right)\right] & \approx\frac{1}{N}\sum_{i}w_{i}g\left(\theta_{i}\right)\\
w_{i} & :=\frac{\pthetapost\left(\theta\right)}{\origpthetapost\left(\theta\right)}=\frac{p\left(x\vert\theta\right)\prior}{C_{\epsilon}}\frac{C_{0}}{p\left(x\vert\theta\right)\origprior\left(\theta\right)}\\
 & =\exp\left(\log\prior-\log\origprior\left(\theta\right)+\log C_{0}-\log C_{\epsilon}\right)\\
C_{0} & :=\int p\left(x\vert\theta\right)\origprior\left(\theta\right)d\theta\\
C_{\epsilon} & :=\int p\left(x\vert\theta\right)\prior d\theta.
\end{align*}

Differentiating the weights,

\begin{align*}
\frac{dw_{i}}{d\epsilon} & =w_{i}\left(\frac{d\log p\left(\theta\vert\epsilon\right)}{d\epsilon}-\frac{d\log C_{\epsilon}}{d\epsilon}\right)\\
 & =w_{i}\left(\frac{d\log p\left(\theta\vert\epsilon\right)}{d\epsilon}-\frac{1}{C_{\epsilon}}\int p\left(x\vert\theta\right)\prior\frac{d\log\prior}{d\epsilon}d\theta\right)\\
 & =w_{i}\left(\frac{d\log p\left(\theta\vert\epsilon\right)}{d\epsilon}-\mbe_{\pthetapost}\left[\frac{d\log\prior}{d\epsilon}\right]\right).
\end{align*}

It follows that
\begin{align*}
\left.\frac{d}{d\epsilon}\frac{1}{N}\sum_{i}w_{i}g\left(\theta_{i}\right)\right|_{\epsilon} & =\frac{1}{N}\sum_{i}w_{i}\left(\frac{d\log p\left(\theta_{i}\vert\epsilon\right)}{d\epsilon}-\mbe_{\pthetapost}\left[\frac{d\log\prior}{d\epsilon}\right]\right)g\left(\theta_{i}\right)
\end{align*}

which is precisely the MCMC estimate of the covariance given by \eqref{local_robustness}.
In particular, when $\epsilon=0$, we have
\begin{align*}
\left.\frac{d}{d\epsilon}\frac{1}{N}\sum_{i}w_{i}g\left(\theta_{i}\right)\right|_{\epsilon=0} & =\frac{1}{N}\sum_{i}\left(\frac{\contamprior\left(\theta_{i}\right)}{\origprior\left(\theta_{i}\right)}-\mbe_{\pthetapost}\left[\frac{\contamprior\left(\theta\right)}{\origprior\left(\theta\right)}\right]\right)g\left(\theta_{i}\right)
\end{align*}

and the importance sampling estimate for replacing $\origprior$ with
$\contamprior$ is
\begin{align*}
\left.\sum_{i}w_{i}g\left(\theta_{i}\right)\right|_{\epsilon=1} & =\frac{1}{N}\sum_{i}\left(\frac{C_{0}\contamprior\left(\theta_{i}\right)}{C_{1}\origprior\left(\theta_{i}\right)}-\mbe_{\pthetapost}\left[\frac{\contamprior\left(\theta_{i}\right)}{\origprior\left(\theta_{i}\right)}\right]\right)g\left(\theta_{i}\right)\\
 & =\frac{C_{0}}{C_{1}}\left.\frac{d}{d\epsilon}\frac{1}{N}\sum_{i}w_{i}g\left(\theta_{i}\right)\right|_{\epsilon=0}
\end{align*}

which confirms that the importance sampling estimate is exactly the
Monte Carlo analogue of the relation \eqref{global_local_sens_evidence}.

In general, we do not know $C_{0}$ and $C_{\epsilon}$ and must use
instead
\begin{align*}
\omega_{i} & :=\frac{\prior}{\origprior\left(\theta\right)}\\
\tilde{\omega}_{i} & :=\frac{\omega_{i}}{\sum_{j}\omega_{j}}.
\end{align*}

Then
\begin{align*}
\frac{d\tilde{\omega}_{i}}{d\epsilon} & =\frac{\omega_{i}}{\sum_{j}\omega_{j}}\frac{d\log p\left(\theta_{i}\vert\epsilon\right)}{d\epsilon}-\frac{\omega_{i}}{\left(\sum_{k}\omega_{k}\right)^{2}}\sum_{j}\omega_{j}\frac{d\log p\left(\theta_{j}\vert\epsilon\right)}{d\epsilon}\\
 & =\frac{\omega_{i}}{\sum_{j}\omega_{j}}\left(\frac{d\log p\left(\theta_{i}\vert\epsilon\right)}{d\epsilon}-\sum_{j}\frac{\omega_{j}}{\sum_{k}\omega_{k}}\frac{d\log p\left(\theta_{j}\vert\epsilon\right)}{d\epsilon}\right)
\end{align*}

which simply replaces the (possibly intractable) expectation $\mbe_{\pthetapost}\left[\frac{d\log\prior}{d\epsilon}\right]$
with its MCMC estimate.

\section{Variational Bayes importance sampling\label{sec:app_Importance-sampling}}

To evaluate \eqref{lrvb_epsilon_sensitivity} requires approximate
integration for which we using importance sampling:
\begin{align*}
\theta_{s} & \sim u\left(\theta\right),\textrm{ for }s=1:S\\
w_{s} & :=\frac{p_{c}\left(\theta_{s}\right)}{u\left(\theta_{s}\right)}\\
\int\frac{\origqthetapost\left(\theta\right)}{\origprior\left(\theta\right)}q_{\eta}\left(\theta\right)^{T}H^{-1}g_{\eta}p_{c}\left(\theta\right)d\theta & \approx\frac{1}{S}\sum_{s=1}^{S}\frac{\origqthetapost\left(\theta_{s}\right)}{p_{0}\left(\theta_{s}\right)}q_{\eta}\left(\theta_{s}\right)^{T}H^{-1}g_{\eta}w_{s}\\
 & =\frac{1}{S}\sum_{s=1}^{S}I_{0}^{q}\left(\theta_{s}\right)w_{s}.
\end{align*}

Note that the influence function can be evaluated once for a large
number of draws from $u\left(\theta\right)$, and then the weights
and prior density can be quickly calculated for any perturbation $\contamprior\left(\theta\right)$,
allowing for fast computation of sensitivity to any $\contamprior\left(\theta\right)$
with little additional overhead.

Since the influence function is mostly concentrated around $\origqthetapost\left(\theta\right)$,
we set $u\left(\theta\right)$ to be $\origqthetapost\left(\theta\right)$
but with quadrupled variance (so that standard deviations are doubled).
Note that this choice of $u$ is a poor approximation of $p_{c}$,
which is nominally the target distribution for importance sampling.
However, since $\origqinfluence$ is very small far from $\origqthetapost\left(\theta\right)$,
it is an adequate approximation of the integral \eqref{lrvb_epsilon_sensitivity}.
Formally, suppose that $\origqinfluence$ is concentrated on a set
$A$ in the sense that $\sup_{\theta\in A^{c}}\left|\origqinfluence\right|\le\epsilon$
for some small $\delta$. Then the absolute error in evaluating \eqref{lrvb_epsilon_sensitivity}
on the set $A$ only is also bounded by $\delta$:

\begin{align*}
\left|\int\origqinfluence p_{c}\left(\theta\right)d\theta-\int_{A}\origqinfluence p_{c}\left(\theta\right)d\theta\right| & =\left|\int_{A^{c}}\origqinfluence p_{c}\left(\theta\right)d\theta\right|\\
 & \le\int_{A^{c}}\left|\origqinfluence\right|p_{c}\left(\theta\right)d\theta\\
 & \le\sup_{\theta\in A^{c}}\left|\origqinfluence\right|\int_{A^{c}}p_{c}\left(\theta\right)d\theta\\
 & \le\delta.
\end{align*}

As long as $\qthetapost\left(\theta\right)$ is chosen to have lighter
tails than $\origprior\left(\theta\right)$ (which is determined by
$\mathcal{Q}$), $\origqinfluence$ will decay quickly away from $\origqthetapost\left(\theta\right)$,
and we can choose $A$ centered on $\origqthetapost\left(\theta\right)$.
Consequently, we can think of $u$ as approximating $\contamprior\left(\theta\right)1_{A}$
rather than $p_{c}\left(\theta\right)$.

\section{Microcredit model\label{sec:app_Microcredit-model}}

We simulate data using a variant of the analysis performed in \citep{meager:2015:microcredit},
though with somewhat different prior choices. In \citet{meager:2015:microcredit},
randomized controlled trials were run in seven different sites to
try to measure the effect of access to microcredit on various measures
of business success. Each trial was found to lack power individually
for various reasons, so there could be some benefit to pooling the
results in a simple hierarchical model. For the purposes of demonstrating
robust Bayes techniques with VB, we will focus on the simpler of the
two models in \citep{meager:2015:microcredit} and ignore covariate
information.

We will index sites with $k=1,..,K$ (here, $K=\numGroups$) and business
within a site by $i=1,...,N_{k}$. The total number of observations
was $\sum_{k}N_{k}=\numObs$. In site $k$ and business $i$ we observe
whether the business was randomly selected for increased access to
microcredit, denoted $T_{ik}$, and the profit after intervention,
$y_{ik}$. We follow \citep{rubin:1981:estimation} and assume that
each site has an idiosyncratic average profit, $\mu_{k1}$ and average
improvement in profit, $\mu_{k2}$, due to the intervention. Given
$\mu_{k}$, $\tau_{k}$, and $T_{ik}$, the \_observed profit is assumed
to be generated according to

\begin{eqnarray*}
y_{ik}\vert\mu_{k},\tau_{k},x_{ik},\sigma_{k} & \sim & N\left(\mu_{k}^{T}x_{ik},\sigma_{k}^{2}\right)\\
x_{ik} & := & \left(\begin{array}{c}
1\\
T_{ik}
\end{array}\right).
\end{eqnarray*}

The site effects, $\mu_{k}$, are assumed to come from an overall
pool of effects and may be correlated:

\begin{eqnarray*}
\mu_{k}\vert\mu & \sim & N\left(\mu,C\right)\\
C & := & \left(\begin{array}{cc}
\sigma_{\mu}^{2} & \sigma_{\mu\tau}\\
\sigma_{\mu\tau} & \sigma_{\tau}^{2}
\end{array}\right).
\end{eqnarray*}

The effects $\mu$ and the covariance matrix $V$ are unknown parameters
that require priors. For $\mu$ we simply use a bivariate normal prior.
However, choosing an appropriate prior for a covariance matrix can
be conceptually difficult \citep{barnard:2000:modeling}. Following
the recommended practice of the software package STAN \citep{stan-manual:2015},
we derive a variational model to accommodate the non-conjugate LKJ
prior \citep{lewandowski:2009:lkj}, allowing the user to model the
covariance and marginal variances separately. Specifically, we use

\begin{eqnarray*}
C & =: & SRS\\
S & = & \textrm{Diagonal matrix}\\
R & = & \textrm{Covariance matrix}\\
S_{kk} & = & \sqrt{\textrm{diag}(C)_{k}}
\end{eqnarray*}

We can then put independent priors on the scale of the variances,
$S_{kk}$, and on the covariance matrix, $R$. We model the inverse
of $C$ with a Wishart variational distribution, and use the following
priors:

\begin{eqnarray*}
q\left(C^{-1}\right) & = & \textrm{Wishart}\left(V_{\Lambda},n\right)\\
p_{0}\left(S\right) & = & \prod_{k=1}^{2}p\left(S_{kk}\right)\\
S_{kk}^{2} & \sim & \textrm{InverseGamma}\left(\alpha_{scale},\beta_{scale}\right)\\
\log p_{0}\left(R\right) & = & \left(\eta-1\right)\log|R|+\constant
\end{eqnarray*}

The necessary expectations have closed forms with the Wishart variational
approximation, as derived in \citet{giordano:2016:robust}. In addition,
we put a normal prior on $(\mu,\tau)^{T}$ and an inverse gamma prior
on $\sigma_{k}^{2}$:

\begin{eqnarray}
p_{0}\left(\mu\right) & = & N\left(0,\Lambda^{-1}\right)\label{eq:microcredit_mu_prior}\\
p_{0}\left(\sigma_{k}^{2}\right) & = & \textrm{InverseGamma}\left(\alpha_{\tau},\beta_{\tau}\right)\nonumber 
\end{eqnarray}

The prior parameters used were: 
\begin{eqnarray*}
\Lambda & = & \left(\begin{array}{cc}
\mcPriorMuInfo & 0\\
0 & \mcPriorMuInfo
\end{array}\right)\\
\eta & = & \mcPriorEta\\
\sigma_{k}^{-2} & \sim & \textrm{InverseGamma}(\mcPriorTauAlpha,\mcPriorTauBeta)\\
\alpha_{scale} & = & \mcPriorScaleAlpha\\
\beta_{scale} & = & \mcPriorScaleBeta\\
\alpha_{\tau} & = & \mcPriorTauAlpha\\
\beta_{\tau} & = & \mcPriorTauBeta
\end{eqnarray*}

As seen in \fig{MCMCComparison} , the means in VB and MCMC match
closely.

\begin{knitrout}
\definecolor{shadecolor}{rgb}{0.969, 0.969, 0.969}\color{fgcolor}\begin{figure}[ht!]

{\centering \includegraphics[width=0.49\linewidth,height=0.196\linewidth]{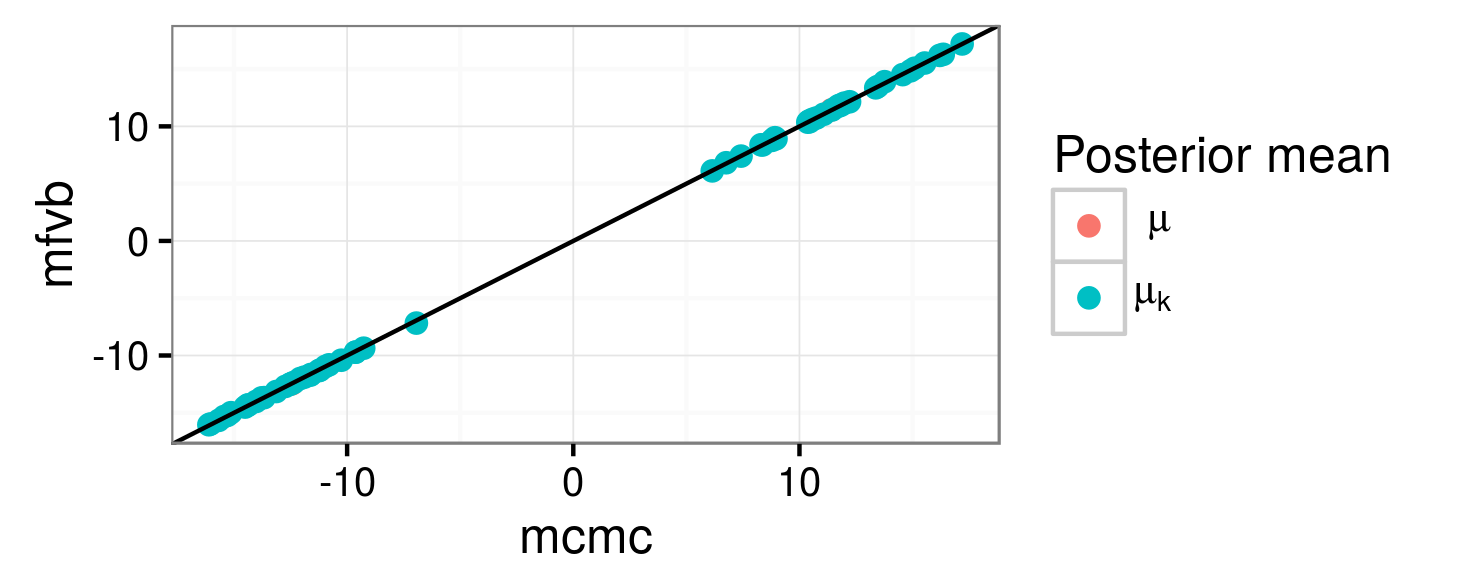} 
\includegraphics[width=0.49\linewidth,height=0.196\linewidth]{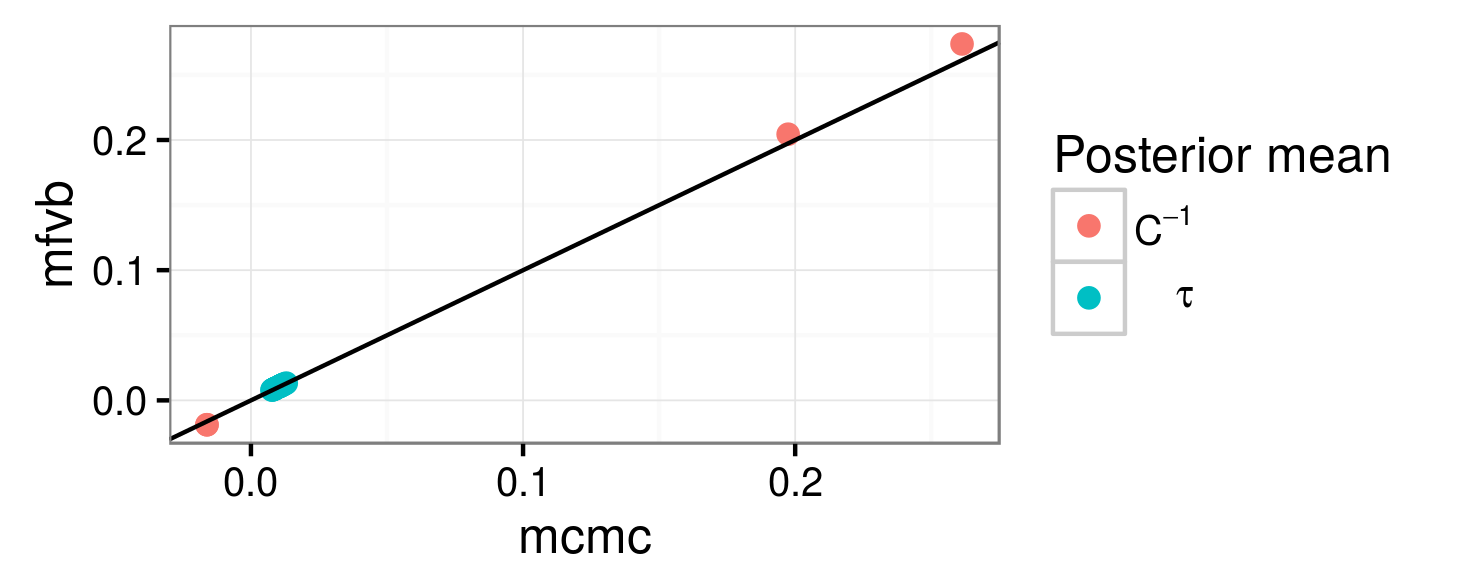} 

}

\caption[Comparison with MCMC]{Comparison with MCMC}\label{fig:MCMCComparison}
\end{figure}

\end{knitrout}

In our simulation, $\mu$ and $\mu_{k}$ are in $\mathbb{R}^{2}$,
so the domain of the prior $p_{0}$$\left(\mu\right)$ is two-dimensional
and $\origqinfluence$ can be easily visualized. We consider the problem
of estimating the effect of replacing the prior on $\mu$ with a product
of independent centered Student t priors. In the notation of \secref{local_robustness},
we take
\begin{align*}
p_{0}\left(\mu\right) & =\mathcal{N}\left(\mu_{1};0,\Lambda^{-1}\right)\cdot\mathcal{N}\left(\mu_{2};0,\Lambda^{-1}\right)\quad\quad\contamprior\left(\theta\right)=\textrm{Student}\left(\mu_{1};\nu\right)\cdot\textrm{Student}\left(\mu_{2};\nu\right).
\end{align*}

We leave all other priors the same, i.e.~$p_{0}\left(\tau_{k}\right)=p_{c}\left(\tau_{k}\right)$
and $p_{0}\left(C\right)=p_{c}\left(C\right)$. In our case, we used
$\nu=\mcPriorStudentTDF$ and $\Lambda=\mcPriorMuInfo$. We will present
sensitivity of $\mbeq\left[\mu_{11}\right]$, the first component
of the first top-level effect. In the notation of \secref{local_robustness},
we are taking $g\left(\theta\right)=\mu_{11}$. Most of the computation
is in generating draws and values for the importance sampling of the
influence function, which can be done once and then reused for any
choice of $\contamprior\left(\theta\right)$ and $g\left(\theta\right)$. 
\end{document}